\documentclass[proceedings]{stacs}
\stacsheading{2010}{155--166}{Nancy}

\usepackage[utf8]{inputenc}
\usepackage[T1]{fontenc}
\usepackage[english]{babel}

\usepackage{amssymb,amscd}
\usepackage{amsfonts}
\usepackage{amsmath}
\usepackage{xspace}
\usepackage{enumerate}
\usepackage{stmaryrd}
\usepackage{color}
\usepackage{graphicx}
\usepackage{subfigure}

\renewcommand{\phi}{\varphi}
\renewcommand{\epsilon}{\varepsilon}
\renewcommand{\Lambda}{\varLambda}
\newcommand{\uz}{\dinf0}

\newcommand{\start}[1]{\begin{#1}\hfill\begin{itemize}}
\newcommand{\finish}[1]{\end{itemize}\end{#1}}

\newcommand{\ignore}[1]{}

\let\kl\mathcal
\newcommand{\N}{\mathbb N}
\newcommand{\Z}{\mathbb Z}
\newcommand{\R}{\mathbb R}
\newcommand{\p}{\kl P}

\let\equi\Leftrightarrow
\newcommand{\tG}{{\tilde G}}
\newcommand{\az}{A^\Z}
\newcommand{\an}{A^\N}
\newcommand{\bz}{B^\Z}
\newcommand{\lang}{\kl L}
\newcommand{\orb}{\kl{O}}

\newcommand{\ktau}{\stackrel\circ\tau}
\newcommand{\deux}{\{0,1\}}
\newcommand{\sett}[2]{\left\{\left.#1\vphantom{#2}\right|#2\,\right\}}
\newcommand{\set}[3]{\sett{#1\in#2}{#3}}
\newcommand{\cc}[2]{\left[#1,#2\right]}
\newcommand{\co}[2]{\left[#1,#2\right[}

\newcommand{\oc}[2]{\left]#1,#2\right]}
\newcommand{\scc}[2]{_{\cc{#1}{#2}}}
\newcommand{\sco}[2]{_{\co{#1}{#2}}}
\newcommand{\soc}[2]{_{\oc{#1}{#2}}}
\newcommand{\sci}[1]{_{\left[#1,+\infty\right[}}
\newcommand{\sio}[1]{_{\left]-\infty,#1\right[}}

\newcommand{\both}[1]{\left\{\everymath{\displaystyle\everymath{}}\begin{array}{l}#1\end{array}\right.}
\newcommand{\soit}[1]{\left|\everymath{\displaystyle\everymath{}}\begin{array}{ll}#1\end{array}\right.}
\newcommand{\appl}[5]{#1:\begin{array}{rcl}#2&\to&#3\\#4&\mapsto&#5\end{array}}
\newcommand{\ou}{\textrm{ where }}
\newcommand{\et}{\textrm{ and }}

\newcommand{\si}{\textrm{ if }}
\newcommand{\sinon}{\textrm{ otherwise}}
\newcommand{\mm}[1]{\textrm{ #1 }}

\newcommand{\restr}[1]{_{\left|#1\right.}}
\newcommand{\length}[1]{\left|#1\right|}

\newcommand{\ipart}[1]{\left\lfloor #1\right\rfloor}
\newcommand{\spart}[1]{\left\lceil #1\right\rceil}
\newcommand{\rev}[1]{\overline{#1}}
\newcommand{\rot}{\gamma}
\newcommand{\cl}[1]{\overline{#1}}
\newcommand{\pow}[1]{^{(#1)}}

\newcommand{\compl}[1]{#1^C}
\newcommand{\eg}{e.g.\@\xspace}
\newcommand{\ie}{\emph{i.e.}\@\xspace}
\newcommand{\etc}{\emph{etc.}\@\xspace}
\newcommand{\resp}[1]{\ [resp. #1]}

\newcommand{\fac}{\sqsubset}
\newcommand{\nfac}{\not\sqsupset}
\DeclareMathOperator*{\id}{id}
\newcommand{\pinf}[1]{\vphantom{#1}^\infty{#1}}
\newcommand{\dinf}[1]{\vphantom{#1}^\infty{#1}^\infty}
\newcommand{\uinf}[1]{#1^\infty}

\newcommand{\pb}[2]{\begin{quote}
{\normalfont\textbf{Instance:}} #1.\\
{\normalfont\textbf{Question:}} #2?
\end{quote}}

\newcommand{\gh}[1]{\boxplus_{#1}}
\newcommand{\dg}[1]{\boxtimes_{#1}}

\begin{document}

\title{Ultimate Traces of Cellular Automata}

\author[upe]{J. Cervelle}{Julien Cervelle}
\address[upe]{Universit\'e Paris-Est, LACL, EA 4219\\
61 Av du G\'en\'eral de Gaulle, 94010 Cr\'eteil Cedex, France}
\email{julien.cervelle@univ-paris-est.fr}

\author[i3s]{E. Formenti}{Enrico Formenti}
\address[i3s]{Laboratoire I3S, Université de Nice-Sophia Antipolis\\
2000, Rte des Lucioles - Les Algorithmes - b\^at. Euclide B - BP 121, 06903 Sophia Antipolis Cedex, France}
\email{enrico.formenti@unice.fr}

\author[cmm]{P. Guillon}{Pierre Guillon}
\address[cmm]{DIM - CMM, UMI CNRS 2807, Universidad de Chile\\
Av. Blanco Encalada 2120,
8370459 Santiago, Chile}
\email{pguillon@dim.uchile.cl}

\thanks{Thanks to the Projet Blanc ANR {\it EMC} and the Comité {\it ECOS-Sud}}


\begin{abstract}
  \noindent A cellular automaton (CA) is a parallel synchronous computing model, which consists in a juxtaposition of finite automata (cells) whose state evolves according to that of their neighbors.
Its trace is the set of infinite words representing the sequence of states taken by some particular cell.
In this paper we study the ultimate trace of CA and partial CA (a CA restricted to a particular subshift). The ultimate trace is the trace observed after a long time run of the CA.
We give sufficient conditions for a set of infinite words to be the trace of some CA and prove the undecidability of all properties over traces that are stable by ultimate coincidence.
\end{abstract}

\maketitle

\section*{Introduction}
Cellular automata are a formal computing model known to display many different dynamical behaviors, from the most simple like nilpotency or equicontinuity to the more complex ones like transitivity, mixing or expansivity. These different behaviors together with their ability to capture many features of natural phenomena increase their popularity in the computer sciencists, mathematicians and physicians communities.

A cellular automaton consists in finite state automata (cells) distributed on a regular lattice (or more generally, on any graph). Each cell updates its state depending on the states of a fixed finite number of neighboring cells. This dependency is given by a local rule which is common to all cells. 

In this paper, we resume our study of traces of cellular automata, that is to say the sequence of states taken by one particular cell.
The main motivation for this work is to study the way scientists deduce general laws from experiments. They proceed by making
experimental observations using a finite number of observation variables (\ie a trace in the context of CA). From these observations,
they conjecture the mathematical law that rules the whole phenomenon. If this  law is verified by (almost all) observations, 
then the scientist concludes that this is the way the phenomenon 
behaves, until  contradicted by new experiments.

However, one also needs formal results ensuring the correctness
of the procedure. Indeed, can any observed trace be generated by a CA? How ``large'' should a trace be to ensure correct reconstruction
of the CA local rule?


The notion of trace for a CA has been studied in~\cite{trace,trice}. In this paper, we proceed with two generalizations: partial traces and ultimate traces. 
A partial trace is the trace of a CA restricted to a particular subshift. This kind of trace is motivated by the fact that there are some experiments where not all initial configurations are admissible: some local constraints have to be respected  (\eg a sand grain cannot be above
an empty cell or two positively charged particles cannot be too close to one another \etc). The ultimate trace is the trace for the long term
behavior \ie when the transient part of the phenomenon is neglected, which is often the case in experimental sciences.

The notion of trace is strictly connected with the concept of symbolic factor. Recall that given a CA $(A^{\Z},F)$, the system $(B^{\N},G)$ is
a (symbolic) \emph{factor} of $(A^{\Z},F)$, if there exists a continuous
surjection $\phi:A^{\Z}\to B^{\N}$ such that $\phi\circ F=G\circ\phi$.
Studying the dynamics of factors is often simpler than studying the
original system. Indeed, traces are special cases of factor systems.
They were introduced as a form of ``back-ingeneering'' tool to lift properties of factors to CA. Along this research direction,
in Section~\ref{sec:undec}, we prove a Rice's theorem for traces.
This is an improvement of a similar result in~\cite{trice},
in the sense that it is more ``natural'' and covers more properties
than the previous one.
\smallskip

The paper is organized into three parts. Section~\ref{sec:defs} recalls main definitions concerning cellular automata and symbolic dynamics.
Sections~\ref{sec:traceability} to~\ref{s:trlim} concern new results about traces. Section~\ref{sec:undec} presents a Rice-like theorem for traces.

\section{Definitions}\label{sec:defs}

Let $\id$ denote the identity map. If $F$ is a function on a set $X$, denote $F\restr Y$ its restriction to some subset $Y\subset X$. If $F$ and $G$ are functions on sets $X$ and $Y$, then $F\times G$ will denote the function on the cartesian product $X\times Y$ which maps any $(x,y)$ to $(f(x),g(y))$.

\paragraph{\em Configurations.}
A \emph{configuration} is a bi-infinite sequence of letters, that is an element of $\az$. The set $\az$ of configurations is the \emph{phase space}.
For integers $i$, $j$, denote $\cc ij$ the set $\{i,\ldots,j\}$, $\co ij$ the set $\cc i {j-1}$, etc\ldots 
For $x\in\az$ and $I=\{i_0,\ldots,i_k\} \subset \N$, $i_0<\cdots<i_k$, note $x_I= x_{i_0}\ldots x_{i_k}$. Moreover, for a word $u$, we note $u\fac x$ if $u$ is a factor of $x$, that is if there exists $i$ and $j$ such that $u=x\scc ij$.
If $u\in A^+$, $\length u$ denotes its length, and $x=\uinf{u}$ \resp{$x=\dinf{u}$} is the infinite word \resp{configuration} such that $x\sco{i\length u}{(i+1)\length u}=u$ for any $i$ in $\N$ \resp{$\Z$}.
A word or a configuration is \emph{uniform} if it is made of a single repeated letter.
If $L\subset A^k$ and $k\in\N\setminus\{0\}$, we shall also note $\dinf L$ the set of configurations $x$ such that $x\sco{ik}{i(k+1)}$ is in $L$ for all $i\in\Z$.
Note that we shall assimilate the sets $\az\times\bz$ and $(A\times B)^\Z$, for alphabets $A,B$.

\paragraph{\em Topology.}
We endow the phase space with the \emph{Cantor topology}.
A base for open sets is given by cylinders: 
for $j,k\in\N$ and a finite set $W$ of words of length $j$, we will 
note $[W]_k$ the \emph{cylinder} $\set w\az{w\sco k{k+j}\in W}$.
$\compl{[W]_k}$ is the complement of $[W]_k$.

\paragraph{\em Cellular automata.}
A (one-dimensional) \emph{cellular automaton} is a parallel
synchronous computation model $(A,m,d,f)$ consisting of cells distributed over a regular lattice indexed by $\Z$. 
Each cell $i\in\Z$ has a state $x_i$ in the finite alphabet $A$, which evolves depending on the state of their neighbors $x\sco{i-m}{i-m+d}$ according to the \emph{local rule} $f:A^d\to A$. The integers $m\in\Z$ and $d>0$ are the \emph{anchor} and the \emph{diameter} of the CA, respectively.
If the anchor is nonnegative, then it can be considered to be $0$ and the automaton is said to be \emph{one-sided}. In this case, a cell is only updated according to its state and the ones of its right neighbors.
The \emph{global function} of the CA (or simply the CA) is $F:\az\to\az$ such that $F(x)_i=f(x\sco{i-m}{i-m+d})$ for every $x\in\az$ and $i\in\Z$.
The \emph{space-time diagram} of initial configuration $x\in\az$ is the sequence of the configurations $(F^j(x))_{j\in\N}$.
When the neighborhood of the CA is symmetrical, instead of speaking of anchor and diameter, we shall simply give a \emph{radius}. A CA of radius $r\in\N\setminus\{0\}$, has $r$ for anchor and $2r+1$ for diameter.

\paragraph{\em Shifts and subshifts.}
The \emph{twosided shift} \resp{\emph{onesided shift}}, denoted $\sigma$, is a particular CA global function defined by $\sigma(x)_i=x_{i+1}$ for every $x\in\az$ and $i\in\Z$ \resp{$x\in\an$ and $i\in\N$} .
According to the Hedlund theorem~\cite{hedlund69}, the global functions of CA are exactly the continuous
self-maps of $\az$ commuting with the twosided shift.

A \emph{twosided subshift} $\Sigma$ is a closed subset of $\az$ with $\sigma(\Sigma)=\Sigma$. A \emph{onesided subshift} $\Sigma$ is a closed subset of $\an$ with $\sigma(\Sigma)\subset\Sigma$.
We simply speak about the \emph{shift} or \emph{subshifts} when the context allows to understand if it is twosided or onesided.

The language of $\Sigma$ is $\lang(\Sigma)=\set w{A^*}{\exists z\in\Sigma,w\fac z}$ and characterizes $\Sigma$, since $\Sigma=\set z{A^\N}{\forall w\fac z,w\in\lang(\Sigma)}$.
For $k\in\N$, denote $\lang_k(\Sigma)=\lang(\Sigma)\cap A^k$.

A subshift $\Sigma$ is \emph{sofic} if $\lang(\Sigma)$ is a
regular language, or equivalently if $\Sigma$ is the set of labels of infinite paths in some edge-labeled graph. In this case, such a graph is called a \emph{graph of $\Sigma$}.

A subshift is characterized by its language 
$\mathcal{F}\subset A^*$ of \emph{forbidden words}, 
\ie such that $\Sigma=\set z\an{\forall u\in\mathcal F,u\nfac z}$. 
A subshift is of \emph{finite type} (SFT for short) if its
language of forbidden words is finite. It is a $k$-SFT (for $k\in\N$) if 
it has  a set of forbidden words of length $k$.
For $\Sigma\subset\az$, define $\orb_\sigma(\Sigma)=\bigcup_{i\in\Z}\sigma^i(\Sigma)$.

\paragraph{\em Partial cellular automata.}
A \emph{partial CA} is the restriction of some CA to some twosided subshift. 

\paragraph{\em Subshift projections.}
If $B\subset A^k$ is an alphabet and $0\leq q<k$, then the $q^\mathrm{th}$ \emph{projection} of an infinite word $x\in B^\N$ is noted $\pi_q(x)\in \an$ and defined by $\pi_q(x)_j=a_q$ when $x_j=(a_0,\ldots,a_{k-1})$.
If $\Sigma$ is a subshift on $B$, we also note $\pi(\Sigma)=\bigcup_{0\le q<k}\pi_q(\Sigma)$, which is a subshift on $A$.
\section{Tracebility}\label{sec:traceability}
\begin{definition}[Traceability]
A subshift $\Sigma\subset\an$ is \emph{traceable} if there exists a CA $F$ on alphabet $A$ whose \emph{trace} $\tau_F = \sett{(F^j(x)_0)_{j\in\N}}{x\in\az}$ is $\Sigma$. In this case, we say that $F$ \emph{traces} $\Sigma$. If $F$ can be computed effectively from data $D$, we say that $\Sigma$ is traceable \emph{effectively from $D$}. In this notion, $D$ can be any mathematical objet, possibly infinite, provided it has a finite representation (SFT, sofic subshifts, regular languages, CA). In this case, it means one of these representations.
\end{definition}

\paragraph{\em Deterministic subshifts.}
Given some $\xi:A\to A$, we call \emph{deterministic subshift} 
the subshift $\orb_\xi=\set{(\xi^j(a))_{j\in\N}}{\an}{a\in A}$.
The following proposition comes from an easy remark on the evolution of uniform configurations -- see Example~\ref{x:110} for a subshift which is not traceable.
\begin{proposition}[\cite{trace}]\label{p:t0}
Any traceable subshift $\Sigma\subset\an$ contains a deterministic subshift $\orb_\xi$ for some $\xi:A\to A$.
\end{proposition}

\paragraph{\em Nilpotent subshifts.}
A subshift $\Sigma\subset\an$ is \emph{$0$-nilpotent} (or simply \emph{nilpotent}) if $0\in A$ and there is some $j\in\N$ such that $\sigma^j(\Sigma)$ is the singleton $\{\uinf0\}$.
It is \emph{weakly nilpotent} if there is some state $0\in A$ such that for every infinite word $z\in\Sigma$, there is some $j\in\N$ such that $\sigma^j(z)=\uinf0$.
Note that a sofic subshift is weakly nilpotent if and only if it admits a unique periodic infinite word, which is uniform.

The following gives another necessary condition for being the trace of a CA.
\begin{theorem}[\cite{nilpeng}]\label{t:fnilptr}
A traceable subshift cannot be weakly nilpotent without being nilpotent.
\end{theorem}

\paragraph{\em Polytraceability.}
When performing some ``back-engineering'' from a trace over an alphabet $A$, \ie when trying to deduce from the trace which CA could
have produced it, it is sometimes easier to design a CA over an
alphabet $B\subseteq A^k$ (for some integer $k$).
Being stacked one atop the other, letters of $B$ can be seen as columns of letters of $A$. In the constructions, the first column is used to produce \emph{all} the elements of $\Sigma$ and the other columns are used to store elements that help to simulate all possible paths along some graph of $\Sigma$. This idea leads to the following notion.

\begin{definition}[Polytraceabilty]
A subshift $\Sigma\subset\an$ is \emph{polytraceable} if there exists a CA $F$ of anchor $0$ and diameter $2$ on alphabet $B\subset A^k$ for some $k$ whose \emph{polytrace} $\ktau_F = \bigcup_{0\leq i<k} \pi_i(\tau_F)$ is $\Sigma$. In this case, we say that $F$ \emph{polytraces} $\Sigma$.
If, furthermore, $B=A^k$, we say that the subshift is \emph{totally polytraceable}. If $F$ and $B$ can be computed effectively from data $D$, we say that $\Sigma$ is (totally) polytraceable \emph{effectively from $D$}.
\end{definition}

Note that a polytrace cannot be weakly nilpotent without being nilpotent, otherwise it would then be the case of the corresponding trace. On the other hand, it need not contain a deterministic subshift.

\begin{theorem}[\cite{trace}]\label{t:sftt3polytr}
 Any subshift $\Sigma$ which is either of finite type or sofic uncountable is polytraceable effectively from $\Sigma$.
\end{theorem}

\paragraph{\em CDD subshifts.}
A sufficient condition for traceability can be given with the help of the following definition.
A subshift $\Sigma\subset\an$ has \emph{cycle distinct from deterministic} property (CDD) if it contains some deterministic subshift $\orb_\xi$ and some periodic infinite word $\uinf w$ such that $w$ contains one letter not in $\xi(A)$. We say that $\Sigma$ is a CDD subshift.

\begin{lemma}[\cite{trace}]\label{l:bordtr2}
Let $\xi:A\to A$, and $\Sigma\subset\an$ be a polytraceable subshift containing a periodic word $\uinf w$, with $w\in A^+\setminus\xi(A)^+$.
Then $\Sigma\cup\orb_\xi$ is traceable effectively from $\xi$, $w$ and a CA polytracing $\Sigma$.
\end{lemma}

This lemma, together with Theorem~\ref{t:sftt3polytr}, gives the following result.
\begin{theorem}[\cite{trace}]\label{t:cddtr}
Any CDD subshift which is either of finite type or sofic uncountable is traceable effectively from the subshift.
\end{theorem}

\section{Partial traceability}\label{s:dg}
We already discussed about partial traceability in the introduction.
Here is the formal definition.
\begin{definition}[Partial traceability]
A subshift $\Sigma$ is \emph{partially traceable} if there exists a partial CA $F$ on an SFT $\Gamma$ whose \emph{trace} $\tau_F = \sett{(F^j(x)_0)_{j\in\N}}{x\in\Gamma}$ is $\Sigma$. In this case, we say that $F$ \emph{partially traces} (or simply \emph{traces}) 
$\Sigma$. If $F$ and some graph of $\Gamma$ can be computed effectively from data $D$, we say that $\Sigma$ is partially traceable \emph{effectively from $D$}.
\end{definition}

Assume that $\Sigma$ is polytraced by some CA $G:\bz\to\bz$, with $B\subset A^h$ and $h\in\N\setminus\{0\}$ -- for instance obtained from Theorem~\ref{t:sftt3polytr}. We simulate it by a partial CA $F$ on some SFT $\Lambda$ in order to get a partial trace instead of a polytrace. This is a kind of \emph{ungrouping} operation that splits \emph{macrocells} 
(on $B$) into independent cells (on $A$).

\paragraph{\em Ungrouping.}
The \emph{ungrouping} operation represents a standard encoding of configurations of $\bz$, with $B\subset A^h$ and $h\in\N\setminus\{0\}$, into configurations of $\az$ and it is defined
as follows
\[\appl{\gh h}{B^\Z}{A^\Z}x{y\text{ such that }\forall i\in\Z,y\sco{hi}{h(i+1)}=x_i~.}\]
We need to be able to perform this encoding locally, we add some constraints to the alphabet $B$.
Indeed, define the twosided subshift $\Lambda=\orb_\sigma(\gh h(\bz))=\bigcup_{0\leq i<h} \sigma^i(\gh h(\bz))$.
We want this union to be disjoint, in order to know, for any configuration of $\Lambda$, up to which shift it can be considered a sequence of macrocells. For this purpose, we add a
\emph{freezing} condition to $B$ as follows.
\paragraph{\em Freezingness.}
A set $W\subset A^h$ is \emph{$p$-freezing}, with $p,h\in\N$, if $\forall i\in\cc1p,A^iW\cap WA^i=\emptyset$, \ie words from $W$ cannot overlap on $h-p$ letters or more.

When $p$ is sufficiently large, we obtain the following property.
\begin{proposition}\label{p:gel2}
Let $W\subset A^h$ be $\ipart{\frac h2}$-freezing, with $h\in\N$. Then $W^2$ is $(h-1)$-freezing; $\Lambda=\bigcup_{0\leq i<h} \sigma^i(\gh h(W^\Z))$ is a disjoint union and an SFT.
\end{proposition}

If $G$ is a CA of radius $1$ on alphabet $B\subset A^h$, we can define its \emph{$h$-ungrouped} partial CA ${\dg h}G$ on the subshift $\Lambda=\orb_\sigma(\gh h(\bz))$, of radius $2h-1$ and local rule:
\[\appl f{\lang_{4h-1}(\Lambda)}Aw{g(u^{-1},u^0,u^1)_i\si\both{w\in A^{h-1-i}u^{-1}u^0u^1A^i\\u^{-1},u^0,u^1\in B\\i\in\co0h\enspace.}}\]

\begin{proposition}\label{p:gelptr}
Let $B\subset A^h$ be $\ipart{\frac h2}$-freezing, and $G$ a CA on alphabet $B$, of radius $1$ and local rule $g:B^3\to B$.
Then the ungrouped CA $\dg hG$ is well defined and its trace is $\ktau_G$.
\end{proposition}
\proof
The local rule $f$ as defined above is not ambiguous since the shift $i$ is unique by Proposition~\ref{p:gel2}. By construction, $f(A^{h-1}u^{-1}u^0u^1A^{h-1})=g(u^{-1}u^0u^1)$, hence by a recurrence on $j\in\N$, we see that if $i\in\co0h$ and $x\in\sigma^i(\gh h(\bz))$, then $\forall k\in\Z,\dg h G^j(x)_0=G^j((x\sco{kh-i}{(k+1)h-i})_{i\in\Z})_i$. As a result, $\tau_{\dg hG}=\bigcup_{0\le i<h}\pi_i(\tau_G)$.
\qed

\paragraph{\em Borders.}

The freezing condition is very restrictive, but any alphabet can be modified in such way to satisfy this property, thanks to a suitable juxtaposition to some freezing set of words.
Formally, a \emph{border} for $B\subset A^k$, with $k\in\N\setminus\{0\}$, is a couple $(\Upsilon,\delta_\Upsilon)$, where $\Upsilon\subset A^l$ is $\ipart{\frac{k+l}2}$-freezing, and $\delta_\Upsilon$ is a function from $\Upsilon$ into itself.
From the latter, seen as the local rule, we define the CA $\Delta_\Upsilon:\Upsilon^\Z\to\Upsilon^\Z$ of radius $0$ whose polytrace is $\bigcup_{0\le i<l}\pi_i(\orb_{\delta_\Upsilon})$.

Borders will be used to separate words representing letters of $B$ in an non-ambiguous way.
\begin{proposition}\label{p:bordtr}
  Let $G$ be a CA on alphabet $B\subset A^k$ and $(\Upsilon\subset A^l,\delta_\Upsilon)$ a border for $B$.
Then, the ungrouped CA $F=\dg{k+l}(\Delta_\Upsilon\times G)$ on the SFT $\Lambda=\orb_\sigma(\dinf{(\Upsilon B)})$ is well defined and its trace is $\ktau_G\cup\ktau_{\Delta_\Upsilon}$.
\end{proposition}
\proof
 If $\Upsilon\subset A^l$ is $\ipart{\frac{k+l}2}$-freezing, then we can see that so is $\Upsilon B$. 
Hence, Proposition~\ref{p:gelptr} can be applied to $\Delta_\Upsilon\times G$, seen as a CA on alphabet $\Upsilon B$.
\qed

In the following, we describe a first example of borders.
\begin{corollary}\label{c:ptrasi}
Let $\Sigma$ be a polytraceable subshift which contains two distinct uniform infinite words $\uinf0$ and $\uinf1$. Then, $\Sigma$ is partially traceable effectively from a polytracing CA and these two words.
\end{corollary}
\proof
Define $\Upsilon^k_{(0,1)}=\{10^k\}$. Note that $\Upsilon^k_{(0,1)}$ is $k$-freezing so $(\Upsilon^k_{(0,1)},\id)$ is a border.
Applying Proposition~\ref{p:bordtr}, as $\ktau_{\Delta_{\Upsilon^k_{(0,1)}}}=\{\uinf0,\uinf1\}$, we get that $\Sigma$ is partially traceable.
\qed

\paragraph{\em Dynamical borders.}
In the case where the polytraceable subshift does not contain two uniform infinite words, we must find another condition to get a freezing alphabet. Assume it contains some periodic non-uniform infinite word $\uinf u$. We note $\rev u=u_{\length u-1}\ldots u_0$ the reverse of $u$ and ${\rot^i(u)}$ the $i^{\textrm{th}}$ rotation $u\sco i{\length u}u\sco0i$ of $u$, for $0\leq i<\length u$.
Then the following represents a border:
let ${\Upsilon^k_u}=\sett{u_i^{k+3\length u}\rev{\rot^i(u)}\rot^i(u)u_i^{\length u}}{0\le i<\length u}\subset A^{k+6\length u}$, and $\delta_{\Upsilon^k_u}:a^{k+3\length u}v{\rev v}a^{\length u}\mapsto v_1^{k+3\length u}\rot(v){\rev{\rot(v)}}v_1^{\length u}$.
\begin{proposition}[\cite{hal}]\label{p:dbord}
$\Upsilon^k_u$ is $(k+3\length u)$-freezing.
\end{proposition}

\begin{corollary}\label{c:ptrnunsi}
Let $\Sigma$ be a polytraceable subshift which contains a periodic infinite word $\dinf u$ of smallest period $\length u>1$. Then, $\Sigma$ is partially traceable effectively from a polytracing CA and $u$.
\end{corollary}
\proof
It is sufficient to apply Proposition~\ref{p:bordtr} to the border $(\Upsilon^k_u,\delta_{\Upsilon^k_u})$. We can see that $\ktau_{\Delta_{\Upsilon^k_u}}=\orb_\sigma(\uinf u)$, which allows to obtain a CA $F:\Lambda\to\Lambda$ such that $\tau_F=\ktau_G$.
\qed

Actually, the only sofic subshifts which are not concerned by the two previous constructions are the nilpotent ones.
\begin{lemma}[\cite{hal}]\label{l:nilpptr}
 Any nilpotent subshift is partially traceable effectively from the subshift.
\end{lemma}

The following gives an example of subshift which is nilpotent, hence partially traceable, but not traceable.
\begin{exa}[\cite{hal}]\label{x:nilpntr}
No CA traces the subshift $\orb_\sigma((\lambda+1+01+001+21)\uinf0)$.
\end{exa}

Putting things together, we get the following important results.
\begin{proposition}\label{p:trpart}
Any polytraceable sofic subshift is partially traceable effectively from a polytracing CA.
\end{proposition}
\proof
It is known that any sofic subshift $\Sigma$ admits some periodic infinite word $\uinf u$, and that it is unique only if $\Sigma$ is weakly nilpotent. In this case, as the projection of some trace, it is nilpotent by Theorem~\ref{t:fnilptr}, and Lemma~\ref{l:nilpptr} allows to conclude.
If there are several distinct periodic infinite words among which one is non-uniform, then we can apply Corollary~\ref{c:ptrnunsi}; otherwise there are several uniform periodic words and we can apply Corollary~\ref{c:ptrasi}.
\qed
The previous proposition, together with Theorem~\ref{t:sftt3polytr}, gives the following -- note that the SFT are partially traceable directly from the definition.
\begin{corollary}
Any uncountable sofic subshift is partially traceable effectively from it.
\end{corollary}

\section{Ultimate traceability}\label{s:trlim}

In this section we consider traces of CA up to ultimate coincidence, \ie assimilating any two subshifts that are different in only a finite number of cells.

One of the difficulties in making traces (Theorem \ref{t:cddtr}), avoided in partial traces, was to deal with ``invalid'' configurations, not in $\orb_\sigma(\gh h(\bz))$. At location of ``errors'' (\ie sites where a pattern of the configuration is not a pattern of $\gh h(\bz)$), instead of applying the simulating rule, we apply a default rule. However, once one of these rules is chosen, the cell must keep using it forever in order to stay in
the ``right'' subshift.

The possibility of initially altering some cells of the subshift simplifies the problem. Indeed, it allows us to build borders in one round and remove all the ``errors'' in the initial configuration. 
We say that two subshifts $\Gamma$ and $\Sigma$ \emph{ultimately coincide} if there exists some generation $J\in\N$ such that $\sigma^J(\Gamma)=\sigma^J(\Sigma)$.

\begin{definition}[Ultimately traceable]
A subshift $\Sigma$ is \emph{ultimately traceable} if there is a CA $G$ such that $\tau_G$ ultimately coincides with $\Sigma$. 
If $F$ and $J$ can be computed effectively from data $D$, we say that $\Sigma$ is ultimately traceable \emph{effectively from $D$}.
\end{definition}

Note that any ultimately traceable subshift is a subsystem of some traceable subshift, and by Proposition~\ref{p:t0} contains some deterministic subshift, but which may not involve all the letters of the alphabet.
\begin{exa}\label{x:110}
Consider the subshift $\Sigma=\orb_\sigma(\uinf{(001)})$. It is an SFT. It is thus polytraceable, but not ultimately traceable since it does not admit any deterministic subshift.
\end{exa}

The proof of the following proposition can be found in the online version.
\begin{proposition}[\cite{hal}]\label{p:polyctr}
Let $\Sigma\subset\an$ be a totally polytraceable subshift which contains some non-nilpotent deterministic subshift $\orb_\xi$, $\xi:A\to A$.
Then $\Sigma$ is traceable effectively from a polytracing CA and $\xi$.
\end{proposition}
With respect to Lemma~\ref{l:bordtr2} two additional hypotheses -- first, that the subshift is totally polytraceable and, second, that the deterministic subshift is not nilpotent -- help get rid of the complex CDD condition, and therefore to get a more precise result about ultimate traces.

\begin{lemma}\label{l:bak}
If $\Sigma$ is a polytraceable subshift, then there exists a subshift $\tilde \Sigma$ such that $\sigma(\Sigma)=\sigma(\tilde\Sigma)$, totally polytraceable effectively from a polytracing CA.
\end{lemma}
\proof
Let $G$ be a CA polytracing $\Sigma$.
Let $\psi:A^k\to B$ be a projection such that $\psi\restr B=\id$; it can be seen as the local rule of some CA $\Psi$ of radius $0$.
Define $\tilde G=G\Psi$.
By construction, we can see that $\tilde G\restr{B^\Z}=G$ and that $\tilde G((A^k)^\Z)=G(B^\Z)\subset B^\Z$, \ie since the second time step the two traces coincide.
\qed

\begin{proposition}\label{p:dglim}
Let $\Sigma\subset\an$ be a polytraceable sofic subshift that contains some deterministic subshift $\orb_\xi$, with $\xi:A'\to A'$ and $A'\subset A$.
Then $\Sigma$ ultimately coincides with some subshift $\tilde\Sigma$ which is traceable effectively from a polytracing CA, $\Sigma$ and $\xi$.
\end{proposition}
\proof
Let $G$ be a CA on $B\subset A^k$ polytracing $\Sigma$, $k\in\N\setminus\{0\}$.
Should we replace $\Sigma$ by the corresponding $\tilde\Sigma$ of Lemma~\ref{l:bak}, we can assume that $B=A^k$.
\begin{itemize}
\item If $\Sigma$ is weakly nilpotent, then, by Theorem~\ref{t:fnilptr}, it is nilpotent, \ie there is some $J\in\N$ such that  
$\sigma^J(\Sigma)=\{\dinf0\}$, property which can be effectively  
tested from $\Sigma$;
any nilpotent CA has a trace which ultimately coincides.
\item If $\orb_\xi$ is not nilpotent, then Proposition~\ref{p:polyctr} can be applied to build a CA whose trace will be the polytrace of $G$.
\item Suppose $\orb_\xi$ is nilpotent, \ie there is some $J\in\N$ and some state $0\in A$ such that $\xi^J(A')=\{0\}$; we define:
\[\appl{\xi'}AAa{0\enspace.}\]
Since the trace $\tau_\tG$ is not weakly nilpotent, it contains some periodic infinite word $\uinf w$, with $w\in A^+\setminus 0^+=A^+\setminus\xi'(A)^+$.
Hence, we can apply Lemma~\ref{l:bordtr2} to build a CA $\tG:\az\to\az$ such that $\tau_\tG=\ktau_G\cup\orb_{\xi'}$.
As a result, $\sigma(\tau_\tG)=\sigma(\ktau_G)\cup\{\dinf0\}=\sigma(\ktau_G)$.
\qed\end{itemize}

\begin{corollary}
Any SFT containing some deterministic subshift and any uncountable sofic subshift containing some deterministic subshift is ultimately traceable effectively from it.
\end{corollary}

Here is an example of subshift which is not traceable, but ultimately traceable.
\begin{exa}[\cite{trace}]\label{x:ctrex}
The subshift $\Sigma=\{\uinf0,\uinf{(01)},\uinf{(10)}\}$ is an SFT and contains some deterministic subshift, but is not traceable.
\end{exa}

The previous corollary is not an equivalence: there are countable sofic ultimately traceable subshifts which are not SFT.
\begin{exa}[\cite{trace}]\label{x:factptr}
 The subshift $(0^*1+1^*)\uinf0$ is sofic, numerable, of infinite type, but traceable.
\end{exa}

The study of the ultimate trace of some CA $F$ is related to that of the limit trace, that is the set $\bigcap_{j\in\N}\sigma^j(\tau_F)$ of traces of configurations which can appear arbitrarily late. In particular, we can see that a surjective subshift which ultimately coincides with the trace of some CA is its limit trace. If it is sofic, the converse is true.

The bitrace of some CA $F$ is the set of its ``biorbits'': 
\[
\tau^*_F=\sett{(x^j_0)_{j\in\Z}}{\forall j\in\Z,x^j\in\az\et F(x^j)=x^{j+1}}.
\]
We can see that it is the twosided subshift with the same language than the limit trace.
As a consequence, we get the following.
\begin{cor}
Any onesided surjective subshift containing some deterministic subshift which is either of finite type or uncountable sofic is the limit trace of some stable CA.
Any twosided subshift containing some deterministic subshift which is either of finite type or uncountable sofic is the bitrace of some stable CA.
\end{cor}

\section{Undecidability}\label{sec:undec}

Let $F$ be a CA of diameter $d$, anchor $m$, local rule $f$ on alphabet $A$.
 A state $0\in A$ is $0$-\emph{spreading} if $d>1$ and for all $u\in A^d$ such that $0\fac u$, we have $f(u)=0$. The CA $F$ is 
\emph{spreading} if it is $s$-spreading for some $s\in A$.

The CA $F$ is \emph{$0$-nilpotent} (or simply \emph{nilpotent}) if there exists a $J>0$ such that $F^J(\az)={\uz}$.
The proof technique developed in \cite{nilpind} allows to prove the following.
\begin{theorem}\label{t:nilpind}
The problem whether a spreading CA $F$ is nilpotent is undecidable.
\end{theorem}
In the sequel, we use the spreading state to control the evolution of
another CA, generalizing the construction used in \cite{trice}.

Consider two CA $F_1$ and $F_2$ of local rules $f_1$ and $f_2$ on (disjoint) alphabets $A_1$ and $A_2$. 
Without loss of generality, assume that they have the same diameter $d$ and anchor $m$. 
Let $A=A_1\cup A_2$, and $\phi:A\to A_1$ be a projection such that $\phi\restr{A_1}=\id$.
Let $N$ and $N_2$ be two CA with the same diameter $d$ and anchor $m$, local rules $n$, $n_2$, and alphabets $B$ and $B_2\subset B$, with $0\in B_2$ being spreading for $N_2$.
We build the CA $H$ of same diameter $d$ and anchor $m$, alphabet $A\times B$ and local rule: \label{d:pcontr}
\[\appl h{(A\times B)^d}{A\times B}{(a_i,b_i)_{-m\le i<d-m}}{\soit{
(f_2(a),n_2(b))&\si a\in A_2^d\et b\in (B_2\setminus\{0\})^d~,\\
(f_1\circ\phi(a),n(b))&\sinon~.}}\]
Starting from a configuration in $(A_2\times B_2)^\Z$, the CA simulates independently $F_2$ and $N_2$ (first part of the rule) until one $0$ appears; at that moment they both change their rules; this change can happen only once for each cell, since from then the letters of the left component remain in $A_1$; hence the two components simulate $F_1$ and $N$ respectively (second part).

The following notions and lemma will help us understand the dynamics of this CA. A set $U\subset A^k$, with $k\in\N\setminus\{0\}$ is \emph{spreading} if $F([U]_1)\subset[U]_0\cap[U]_1$ or $F([U]_0)\subset[U]_0\cap[U]_1$.
If $F$ is a CA on alphabet $A$ and $A'\subset A$, then we say that $F$ is (globally) \emph{$A'$-mortal} if $\forall x\in\az,\exists i\in\Z,\exists j\in\N,F^j(x)_i\in A'$.
\begin{lemma}\label{l:sprnilpu}
If $F$ is a CA on alphabet $A$ and $A'\subset A$ is spreading, then $F$ is $A'$-mortal if and only if $\exists J\in\N,\forall x\in\az,\forall i\in\Z,\forall j\ge J,F^j(x)_i\in A'$.
\end{lemma}
\proof
Suppose $F$ is $A'$-mortal. By compacity, there is some $J\in\N$ and some radius $I\in\N$ such that $\forall x\in\az,\exists i\in\cc{-I}I,F^J(x)_i\in A'$. If $A'$ is left-spreading, we obtain thanks to a trivial recurrence, $\forall x\in\az,F^{J+2I}(x)_{-I}\in A'$. Thanks to uniformity and shift-invariance, we obtain the stated result. The right-spreading case is symetric.
\qed
\begin{lemma}\label{l:separ2}~\begin{itemize}
\item If $N_2$ is nilpotent, then there is some $J\in\N$ such that $\pi_0( H^J((A\times B)^\Z))\subset A_1^\Z$ and then, on $H^J((A\times B)^\Z)$, $H$ behaves like $F_1\times N$.
\item Otherwise, there is a subshift $\Lambda\subset B_2^\Z$ such that $\pi_0\circ H\restr{A_2^\Z\times\Lambda}=F_2\circ\pi_0$.
\end{itemize}
\end{lemma}
\proof\begin{itemize}
 \item Suppose $N_2$ is nilpotent. From the definition of $H$, no orbit implies always the first part of the rule: $H$ is $A_1\times B$-mortal. Moreover we can see that $A_1\times B$ is spreading for $H$. Thanks to Lemma~\ref{l:sprnilpu}, $H$ remains ultimately on the alphabet $A_1\times B$.
 \item Otherwise, there exists, thanks to Lemma~\ref{l:sprnilpu}, some configuration $x\in B_2^\Z$ such that $\forall i\in\Z,\forall j\in\N,N_2^j(x)_i\ne0$; the subshift $\Lambda=\cl{\orb_\sigma(\orb_N(x))}$ is such that $A_2^\Z\times\Lambda$ is $H$-invariant and its first column is $F_2$.
\qed\end{itemize}
Since they are reduced to the nilpotency of the spreading CA $N_2$, the two cases presented are recursively inseparable, provided that they are disjoint.

\paragraph{\em Properties of ultimate polytraces.}
As for the conditions of traceability, polytraces represent here a useful intermediary tool.

Let $G$ be a CA on alphabet $\deux$ and $N$ a CA on alphabet $\deux$ of radius $0$ and locale rule $\xi:\deux\to\deux$ such that $\orb_\xi\subset\tau_G$. We build the alphabets $A_1=\sett{(a,a,b)}{a,b\in\deux}$ and $A_2=\deux^3\setminus A_1$, as well as the CA $F_1=(N\times N\times G)\restr{A_1}$, $F_2=(\sigma\times\sigma\times G)\restr{A_2}$. We can apply Lemma~\ref{l:separ2} to the CA $H$ built as above from $F_1$, $F_2$, $N$, and any $0$-spreading CA $N_2$ on alphabet $\deux$.

The product is here composed of four layers. The fourth one controls the whole behavior thanks to its spreading state $0$. The third one simulates $G$ independently. When the two first ones are distinct, they simulate full shifts (whose trace is $\deux^\N$) that hide the trace of $G$. As soon as some $0$ appears in the last layer, they stop, unify and then apply $\xi$, which is contained in $\tau_G$.

In the end of the section, we consider that $H$ is built from $G$, $N$ and $N_2$, the CA $F_1$ and $F_2$ being defined as above.

\begin{lemma}\label{l:separ3}~
 \begin{itemize}
  \item If $N_2$ is nilpotent, then $\ktau_H$ ultimately coincides with $\tau_G$.
\item Otherwise, $\ktau_H=\deux^\N$.
 \end{itemize}
\end{lemma}
\proof~\begin{itemize}
\item Thanks to Lemma~\ref{l:separ2}, if $N_2$ is nilpotent, then the first three components of $H$ and $(N\times N\times G)\restr{A_1}$ ultimately coincide, and that of the forth component is ultimately included in $\tau_N$. Considering that the polytrace of $(N\times N\times G)\restr{A_1}$ is $\tau_G \cup \tau_N$ and that, by hypothesis, $\tau_N\subset\tau_G$ the polytrace of $H$ ultimately coincides with $\tau_G$.
\item Otherwise, there exists a subshift $\Lambda$ such that the partial CA $H\restr{A_2^\Z\times\Lambda}$ admits as first three projections $(\sigma\times\sigma\times G)\restr{A_2^\Z}$. The first projection of the trace is $\deux^\N$, since for any infinite word $a$, there is another word $b$ distinct in every cell ($\forall i\in\N$, $a_i\ne b_i$); hence the trace $\tau_H$ contains and therefore is $\deux^\N$.
\qed\end{itemize}

\paragraph{\em Properties of traces.}
As in the previous section, we are now going to simulate CA on alphabets with several components to transform the result on polytraces into a result on traces.

\begin{lemma}\label{l:separ4}
Let $G$ be a non-nilpotent onesided CA whose trace is not $\deux^\N$. The set of CA on alphabet $\deux$ whose trace is $\deux^\N$ is recursively inseparable from the set of CA on alphabet $\deux$ whose trace ultimately coincides with $\tau_G$.
\end{lemma}
\proof Let $N_2$ be a onesided $0$-spreading CA. \begin{itemize}
\item Suppose that the trace $\tau_G$ contains some non-nilpotent deterministic subshift $\orb_\xi$, with $\xi:\deux\to\deux$. $\xi$ can be seen as the local rule of the CA $N$. Build CA $H$ as before.
From Proposition~\ref{p:polyctr}, $H$ can be transformed into some CA $F$ on alphabet $\deux$ such that $\tau_F=\ktau_{H}$.
\item If the trace $\tau_G$  does not contain any non-nilpotent deterministic subshift, then, as it is still non-nilpotent, it contains some periodic infinite word $\uinf w$, $w\in\deux^*$, $w\notin 0^*$. We can define the null CA $N=\overline0$ on $\deux^\N$ of local rule $\xi':a\mapsto0$ and define $H$ as before. Remark that $\uinf w$ and $\uinf0$ are in the trace of $H$, hence we can apply Lemma~\ref{l:bordtr2} to build a CA $F$ on alphabet $\deux$ such that $\tau_F=\ktau_{H}$.
\end{itemize}
In both cases, Lemma~\ref{l:separ3} gives that if $N_2$ is $0$-nilpotent, then $\tau_F$ ultimately coincides with $\tau_G$, otherwise $\tau_F=\deux^\N$.
As $F$ is computable from $G$, were the two cases separable, Theorem~\ref{t:nilpind} would be contradicted.
\qed
From the remark that some CA traces are not equal to the full shift, we can see that this behavior is undecidable. But the previous lemma also infers other nontrivial properties of traces.

A property $\p$ over subshifts is \emph{stable by ultimate coincidence} if for any subshifts $\Sigma$ and $\Gamma$ which ultimately coincide, we have $\Sigma\in\p\Longleftrightarrow\Gamma\in\p$.
\begin{theorem}\label{t:trice}
Let $\p$ be a property over subshifts which:
\begin{enumerate}
 \item is satisfied by the trace subshift of some CA over alphabet $\deux$, but not all;
 \item is stable by ultimate coincidence.
\end{enumerate}
Then, the problem
 \pb{a CA $G$ on alphabet $\deux$}{does $\tau_G$ satisfy property $\p$}
 is undecidable.
\end{theorem}
\proof
Let $\p$ be such a property and assume that $\deux^\N$ does not satisfy $\p$, should we take the complement. If $\p$ is only satisfied by nilpotent subshifts, then thanks to stability by ultimate coincidence, it is equivalent either to $0$-nilpotency, to $1$-nilpotency or to nilpotency, which are all undecidable by Theorem~\ref{t:nilpind}.
Otherwise, $\p$ is satisfied by the trace $\tau_G$ of some non-nilpotent CA $G$. Would an algorithm decide $\p$, it would allow to separate the trace $\tau_G$ to $\deux^\N$ among traces over alphabet $\deux$ up to ultimate coincidence, contradicting Lemma~\ref{l:separ4}.
\qed

This result includes in particular the so-called ``nilpotent-stable'' properties defined in \cite{trice}, such as fullness, finiteness, ultimate periodicity, soficness, finite type, inclusion of a particular word as a factor. It also includes nilpotency, as well as all properties of the trace of the limit system $(\bigcap_{J\in\N}F^j(\az),F)$ of CA $F$, as stated in \cite{these}.
Moreover, it can be easily adapted to larger traces, \ie taking the states of a central group of cells of each configuration.
We can also see that this theorem implies the undecidability of all of these properties of any line projection of two-dimensional SFT (tilings respecting local constraints).

\section{Conclusions}
\ignore{
The idea of studying factors of a dynamical system dates back to
H. Poincar\'e and its three body problem. Most of the times
factors have a dynamics easier to understand than the original
system. In the context of CA, factors arise in a very natural manner.
Indeed, consider a CA $F$ and a finite open cover $O$ of same size
as the alphabet of $F$. Then, encoding orbits of $F$ according to the
labels of the sets of $O$ visited gives the \emph{natural factor} of $F$.

Natural factors have proved a valuable tools for the study of CA dynamics~\cite{classif,nasu}. !!!!!!!!!!TERMINER!!!!!!!!!
}

In our study of CA traces, we have reached two kinds of important results. On the one hand, we provided sufficient conditions for a subshift to be a polytrace, a trace, a partial trace, an ultimate trace. 
On the other hand, we proved the undecidability of nearly all properties over ultimate traces. Going beyond undecidability, when
it is clear that the trace has been generated by CA, 
it would be interesting to study which ones, and with which minimal radius. 

Remark that the contructions used in the paper build CA with a very large radius. It would be interesting to study the traces produced by cellular automata of a given fixed radius. This is not a so great limitation in complexity, since
elementary CA (binary alphabet, radius $1$) already present
rich different behaviors. In particular, a deeper study of the so-called ``canonical factors'', \ie traces which width is the radius of the CA, could be fundamental to fully understand this notion.

Another interesting research direction consists in trying to adapt
or find some refinement of
K\r{u}rka's language classification (\cite{classif}) to the case of
traces or ultimate traces. This would provide an interesting link between the complexity of the dynamics of CA and
the (language) complexity of its traces.
\bibliographystyle{alpha}
\bibliography{tra}
\newpage
\appendix

\section{Partial traceability}

Let the trace application of a cellular automaton $F$ on alphabet $A$ be defined by
\[
\appl{T_F}{\az}{\an}{x}{(F^j(x)_0)_{j\in\N}~.}
\]

\proof[Proof of Proposition~\ref{p:gel2}]
 Suppose $W$ is $\ipart{\frac h2}$-freezing. Then so is $W^2$. Moreover, if $\ipart{\frac h2}<i<h$, then $A^iW^2\cap W^2A^i\subset A^i(WA^{h-i}\cap A^{h-i}W)A^i=\emptyset$; hence $W^2$ is $(h-1)$-freezing. Consequently, if $1\le i<h$, then $\gh h(\bz)\cap\sigma^i(\gh h(\bz))\subset[BB]_0\cap[BB]_i=\emptyset$, hence the union is disjoint.
 Now take $x\in\az$ such that $\forall i\in\Z,z\sco i{i+2h}\in\lang(\orb_\sigma(\gh h(\bz)))$. Let us show by recurrence on $k\ge2h$ that $\exists i\in\co0h,\exists y\in\sigma^i(\gh h(\bz)),x\sco0k=y\sco0k$.
The base case exactly corresponds to our hypothesis, in the center.
Suppose $i\in\co0h$, $y\in\sigma^i(\gh h(\bz))$ and $k\ge2h$ are such that $x\sco0k=y\sco0k$. By hypothesis, we know that there exists $i'\in\co0h$, $y'\in\sigma^{i'}(\gh h(\bz))$ such that $x\soc{k-2h}k=y'\soc{k-2h}k$.
We obtain $x\in[B]_{\ipart{\frac{k-1}h}h-i}\cap[B]_{\ipart{\frac{k-1}h}h-i'}$, hence, by freezingness, $i=i'$. Besides, if we define $z\in\sigma^i(\gh h(\bz))$ by $z\sio{\ipart{\frac{k-1}h}h-i}= x\sio{\ipart{\frac{k-1}h}h-i}$ and $z\sci{\ipart{\frac{k-1}h}h-i}= y\sci{\ipart{\frac{k-1}h}h-i}$, then $x\scc0k=z\scc0k$; therefore the property is hereditary.
A similar recurrence on $k<0$ can be used to get, at the limit, that $\exists i\in\co0h,x\in\sigma^i(\gh h(\bz))$.
\qed

\proof[Proof of Proposition~\ref{p:dbord}]
Let $l=k+6\length u$. Suppose we have two words $v= a^p\rev uua^k$ and $w= a'^p\rev{u'}u'a'^k$ such that $\exists i\in\co0p,v\sco il=w\sco0{l-i}$.
Then first note that $a=v_i=w_0=a'$.
\begin{itemize}
\item If $k\le i\le p$, then we have $\rev u=v\sco p{p+k}=w\sco{p-i}{p+k-i}=a^k$, which is forbidden by the definition.
\item If $0\le i<k$, then we can use the symmetry of the words $\rev uu$ and $\rev{u'}u'$.
Let $m=\min\set q{\co0l}{v_q\ne a}=i+\min\set q{\co0l}{w_q\ne a}$. On the one hand:
\begin{equation}\label{eqmin}\begin{array}{lll}
m&=&p+\min\set q{\co0k}{u_q\ne a}\\
&=&p+k-\max\set q{\co0k}{\rev u_q\ne a}\\
&=&2p+k-\max\set q{\co0l}{v_q\ne a}\enspace.
\end{array}\end{equation}
On the other hand:
\begin{equation}\label{eqmax}\begin{array}{lll}
m&=&i+p+\min\set q{\co0k}{u'_q\ne a}\\
&=&i+p+k-\max\set q{\co0k}{\rev{u'}_q\ne a}\\
&=&2i+2p+k-\max\set q{\co0l}{v_q\ne a}\enspace.
\end{array}\end{equation}
It results that $i=0$.
\qed\end{itemize}

\proof[Proof of Lemma~\ref{l:nilpptr}]
 Let $\Sigma\subset\an$ be a subshift and $0\in\Sigma$ such that $\exists J\in\N,\sigma^J(\Sigma)=\{\uinf0\}$.
 Let $B=\sett{0^{3J}\rev uu0^J}{u\in\lang_J(\Sigma)}\subset A^h$, with $h=6J$, and $\Delta_B$ the CA on alphabet $B$, of radius $0$ and local rule:
 \[\appl{\delta_B}BB{0^{3J}\rev uu0^J}{0^{3J+1}\rev{u\scc0{J-2}}u\sco0{J-2}0^{J+1}\enspace.}\]
  First remark that
  \[
 \forall i\in\co{4J}{6J},\pi_i(T_{\Delta_B}(0^{3J}\rev uu0^J))=\sigma^{i-4J}(u\uinf0)
 \]
  and 
  \[
  \forall i\in\co0{4J},\pi_i(T_{\Delta_B}(0^{3J}\rev uu0^J))=\sigma^{4J-1-i}(u\uinf0)\enspace.
  \]
   Globally, $\ktau_{\Delta_B}=\Sigma$.
On the other hand, $C=B\setminus\{0^{6J}\}$ is $3J$-freezing. By Proposition~\ref{p:gelptr}, we can build without ambiguity a partial CA $F$ on the subshift $\Lambda=\orb_\sigma(\gh h(\bz))$, of radius $h-1$ and local rule:
 \[\appl f{\lang_{2h-1}(\Lambda)}Aw{\soit{\delta_B(u)&\si w=A^iuA^{h-1-i},u\in C\et i\in\co0h\\0&\sinon~,}}\]
which statisfies $\tau_F=\ktau_{\Delta_B}=\Sigma$.
\qed

\proof[Proof of Example~\ref{x:nilpntr}]
Suppose $F$ is a CA on alphabet $\{0,1,2\}$ of trace $\tau_F=\orb_\sigma((2+00)1\uinf0)$.
Note that, if $x\in\{1,2\}^\Z$, then for any $i\in\Z$, $F(x)_i=x_i-1$, hence $F(\{1,2\}^\Z)=\{0,1\}^\Z\supset F(\az)$.
As a result $F^2(\az)\subset F^2(\{1,2\}^\Z)=\{\dinf0\}$, which contradicts the fact that $001\uinf0\in\tau_F$.
\qed

\section{Semifinite automata}
Let us build the set of finite and infinite words over some given alphabet $A$: $\overline\az= A^*\sqcup\az$. We index a finite word $u$ from $-\ipart{\frac{\length u}2}$ to $\spart{\frac{\length u}2}$ and we endow that space with the following metric:
\begin{align*}
 \appl d{\overline\az^2}{\R_+}{(x,y)}{2^{-\min\sett{k}{k\in D_{x,y}}}}\\
\ou k\in D_{x,y}\equi\soit{x\scc{-k}k\textrm{ exists, but not }y\scc{-k}k\\
y\scc{-k}k\textrm{ exists, but not }x\scc{-k}k\\
x\scc{-k}k\ne y\scc{-k}k\enspace.}
\end{align*}
In particular, $\overline\az$ is the closure of $A^*$.
A \emph{semifinite automaton} is a dynamical system $F$ over space $\overline\az$ which induces some CA over $\az$ and some subsystem over each $A^k$, $k\in\N$.

Visually we can see a semifinite automaton as a CA on the SFT $\Sigma_{\compl LL+R\compl R}\subset\bz$, where $B=A\sqcup\{L,R\}$; the alphabet is added two still fresh letters $L$ and $R$, such that $aL$ and $Rb$, with $a\ne L$ and $b\ne R$, are forbidden. The configurations which only have $R$ at the right and $L$ at the left stand for the finite words, up to some shift.

By this caracterisation, we can see that there is a diameter $d\in\N$, an anchor $m\in\N$ and a local rule $f:(L^*A^*\cap A^m)A(A^*R^*\cap A^{d-m-1})\to A$.
The trace application $T_F$ is defined only for sufficiently large words: denote $\tau_F=T_F(\set x{\overline\az}{x_0\textrm{ exists}})$. The following proposition shows that such traces can correspond to that of onesided CA.
\begin{proposition}\label{p:acp}
Any onesided CA $G$ on $\az$ can be extended to some semifinite automaton $F$ on space $\overline\az$ with $\tau_G=\tau_F$.
\end{proposition}
\proof
The local rule $f$ of $F$ is equal on $\az$ to $g$, that of $G$. It is thus sufficient to define the function $f$ for the extremities of the finite words, \ie when $L$ or $R$ appear in the neighborhood, in such a way that the trace produced in these cells corresponds to the trace of some configuration in $\az$. It is sufficient to show that $\forall k\in\N,T_F^{\cc{-k}k}(A^{\cc{-k}k})\subset\tau_G^{\cc{-k}k}$; we will then have, by projection, $T_F^{\cc{-l}l}(A^{\cc{-k}k})\subset\tau_G^{\cc{-l}l}$ for every width $l<k$.
$g$ can be extended in:
\[\appl f{A(A\sqcup\{R\})^{d-1}}Au{\soit{g(u)&\si R\nfac u\\g(u\sco0ku_k^{\length u-k})&\si\exists k\in\co0{\length u-1},u_k\in A\et u_{k+1}=R.}}\]
A trivial recurrence shows that the trace of every word $u\in A^{\cc{-k}k}$ is 
\[
T^{\cc{-k}k}_F(u)=T^{\cc{-k}k}_G(\pinf{u_0}u\uinf{u_{\length u}})\subset\tau_G\enspace.
\]
\qed

\section{Total ungrouping}
For $\xi:A\to A$, we define \[\Upsilon^k_\xi=\sett{ab^{k+1}}{\forall j\in\N,\xi^j(a)\ne\xi^j(b)}~,\]
\[\et\appl{\delta_\Upsilon}{\Upsilon^k_\xi}{\Upsilon^k_\xi}{ab^{k+1}}{\xi(a)\xi(b)^{k+1}~.}\]

\begin{lemma}\label{l:gelks0}
 ${\Upsilon^k}=\sett{ab^{k+1}}{a\ne b}\subset a^{k+2}$ is $k$-freezing.
\end{lemma}
\proof
 If $u= ab^{k+1}$ and $v= a'b'^{k+1}$ satisfy $u\scc i{k+1}=v\scc0{k+1-i}$ with $1\le i\le k$, then on the one hand $b=u_i=v_0=a'$ and on the other hand $b=u_{i+1}=v_1=b'$; summing up, we have $a'=b'$, \ie $v$ is not a word from the border.
\qed

For any function $\xi$, we note $\xi\pow k=\xi\times\ldots\times\xi$ ($k$ times).

\begin{lemma}\label{l:gelks}
 If $\xi:A\to A$ is such that $\orb_\xi$ is not nilpotent, then $\Upsilon^k_\xi$ is nonempty.
$(\Upsilon^k_\xi,\xi\pow{k+2})$ is then a border for $A^k$.
\end{lemma}
\proof
If $\xi$ is not nilpotent, it can be seen that there are two letters $a,b\in A$ such that $\forall j\in\N,\xi^j(a)\xi^j(b)$, \ie $ab^{k+1}\in\Upsilon^k_\xi$.
Moreover, $\Upsilon^k_\xi$ is $k$-freezing, as a subset of $\Upsilon^k$\ref{l:gelks0}.
Finally, if $ab^{k+1}$ is a word of the border $\Upsilon^k_\xi$, \ie $\forall j\in\N,\xi^j(a)\ne\xi^j(b)$, then in particular $\forall j\in\N,\xi^j\xi(a)\ne\xi^j\xi(b)$, hence $\xi\pow{k+2}(ab^{k+1})\in\Upsilon^k_\xi$.
\qed

\begin{lemma}\label{l:geltr0}
  Let $\xi:A\to A$ and $B\subset A^{2p}$ be $p$-freezing and such that $\left(\xi\pow{2p}\right)^{-1}(B)\subset B$. Let $\tilde G$ be a semifinite automaton of anchor $0$, diameter $2$, local rule $\tilde g:B(B\cup\lambda)^2\to B$, such that $\pi\sco0p(\tilde G)=\xi\pow p$.
Then we can build a CA $F:\az\to\az$ such that $\tau_F=\ktau_{\tilde G}\cup\orb_\xi$.
\end{lemma}
\proof
$B$ being freezing, we can well define the CA $F$ of diameter $d=10p-1$, anchor $m=2p-1$ and local rule:
\[\appl f{A^d}Aw{\soit{\tilde g(u^0,u^1)_i&\si\both{w\in A^{m-i}u^0u^1u^2u^3A^i,i\in\cc0m,\\
u^0,u^1\in B,u^2u^3\in\Theta_B}\mm{(execution),}\\
\tilde g(u^0,\lambda)_i&\si\both{w\in A^{m-i}u^0u^1u^2A^i,i\in\cc0m,\\
u^0\in B,u^1u^2\in\Theta_B,u^2u^3\notin\Theta_B}\mm{(frontier),}\\
\xi(w_0)&\sinon\textrm{(default).}}}\]
Note the implicit existence of an external frontier, subcase of the default mode: the rightmost of a finite juxtaposition of cells evolves according to $\xi$ in order not to be able to create a new macrocell to its right, in the overlapping zone between the two modes. Now the hypotheses forbid both the execution and the default mode to create a word of $B$ from scratch, thanks to some ``no man's'' macrocell, which evolves in default mode. Let us express this property formally.
\begin{itemize}
 \item Let $x\in[B]_0$; suppose there exists some cell $i\in\co0{2p}$ such that $F(x)_k\ne\xi(x_k)$. The cell $k$ applies the execution or frontier mode, hence $x\in[B\Theta_B]_i$ with $k-2p<i\le k$. The freezing property of $B$ gives that $i$ and $i+2p$ are not in $\cc{-p}p$. Since $i>k-2p\ge-2p$, we obtain $i>p$; therefore the position in $\Theta_B$ of the considered cell $k-i$ is less than $p$ and our hypothesis $F(x)_k\ne\xi(x_k)$ contradict the hypothesis on $\tilde G$.
We showed $F(x)\sco0{2p}=\xi\pow{2p}(x\sco0{2p})\in B$. As a result:
\begin{equation}\label{eq:bb1}
 F([B]_0)\subset[B]_0\enspace.
\end{equation}
\item Now let $x$ a configuration such that $F(x)\in[B]_0$; suppose that there is a cell $k\in\co0{2p}$ such that $F(x)_k\ne\xi(x_k)$. Then there exists $i\in\oc{k-2p}k$ such that $x\in[B\Theta_B]_i$. \eqref{eq:bb1} gives $F(x)\in[BB]_i$, and $B$ being freezing, we obtain $i,i+2p\notin\cc{-p}p$. Since $i>k-2p\ge-2p$, we have $i>p$; hence the position in $\Theta_B$ of the considered cell $k-i$ is less than $p$ and our hypothesis $F(x)_k\ne\xi(x_k)$ contradicts the hypothesis on $\tilde G$.
Similarly to the previous point, we can conclude:
\begin{equation}\label{eq:bb}
 F^{-1}([B]_0)\subset[B]_0\enspace.
\end{equation}
\item The equation \eqref{eq:bb} can be rewritten into $F([\compl B]_0)\subset[\compl B]_0$ and the definition of macrocells into $[\Theta_B]_0=[B]_0\cap\bigcap_{1\le i<2p}[\compl B]_i$. Combining with \eqref{eq:bb1}, we get:
\begin{equation}
 F([\Theta_B]_0)\subset[\Theta_B]_0\enspace.
\end{equation}
\item Now suppose we have a configuration $x$ such that $F(x)\in[\Theta_B]_0$.
We know by \eqref{eq:bb} that $x\in[B]_0$. If on top of that $x$ were in a cylinder $[B]_i$, with $i\in\co0{2p}$, then so would $F(x)$ by \eqref{eq:bb1}, which would contradict our hypothesis. As a result:
\begin{equation}
 F^{-1}([\Theta_B]_0)\subset[\Theta_B]_0\enspace.
\end{equation}
\end{itemize}
We have a partition into three disjoint $F$-invariant sets:
\begin{align*}\az&=X_1\sqcup X_2\sqcup X_3\\
\ou X_1&=\bigcup_{0\le i<2p}\bigcap_{q\in\Z}[B]_{2pq-i}\mm{(valid configurations),}\\
X_2&=\bigcup_{\substack{0\le i<2p\\q>0}}[B\Theta_B]_{-i}\cap\compl{[\Theta_B]_{2pq-i}}\mm{(invalid configurations with valid center),}\\
X_3&=\bigcap_{0\le i<2p}\compl{[B\Theta_B]_i}\mm{(configurations with invalid center).}
\end{align*}
The trace can be decomposed the same way:
\[\tau_F=T_F(X_1)\cup T_F(X_2)\cup T_F(X_3)\enspace.\]
Remark that $X_1=\orb_\sigma(\gh h(\bz))$ and that $F\restr{X_1}$ is the partial CA defined by Proposition~\ref{p:gelptr}; consequently its contribution $T_F(X_1)$ to the trace is the polytrace $\ktau_{\tilde G\restr\bz}$ of the corresponding CA.
From the third part of the definition, the default mode on the central cell only consists in applying $\xi$; therefore, $T_F(X_3)=\sett{(\xi^j(a))_{j\in\N}}{a\in A}$.
Now let $x\in X_2$, \ie $\exists i\in\co0{2p},q>0,x\sco{-i}{2p(q+2)-i}=u^0\ldots u^qu^{q+1}u^{q+2}$, with $u^su^{s+1}\in\Theta_B$ for $0\le s\le q$ and $u^{q+1}u^{q+2}\notin\Theta_B$. We can see $u=u^0\ldots u^{q-1}$ as a finite word of $B^+$.
The definition and a trivial recurrence give that $\forall j\in\N,F^j(x)\sco{2p(q+1)-i}{2p(q+3)-i}\notin\Theta_B$ and $\forall s\in\cc0q,F^j(x)\sco{2ps-i}{2p(s+2)-i}\in\Theta_B$ and $\forall s\in\co0q,F^j(x)\sco{2ps-i}{2p(s+1)-i}=\tilde G^j(u)_s$.
As a result, $T_F(x)=\pi_i(T_{\tilde G}(u))$, hence $T_F(X_2)=\ktau_{\tilde G\restr{B^+}}$.
Putting things together, we obtain $\tau_F=\ktau_{\tilde G}\cup\orb_\xi$.
\qed

\begin{lemma}\label{l:bordtr0}
  Let $\xi:A\to A$, $G$ be a onesided CA on alphabet $A^k$ and $(\Upsilon\subset A^l,\xi\pow l)$ a border for $A^k$, with $l>k>0$.
Then we can build a CA $F:\az\to\az$ such that $\tau_F=\ktau_G\cup\orb_\xi$.
\end{lemma}
\proof
Should we take a larger $k$, we can assume the radius of $G$ is $1$. Let $\Delta_\Upsilon$ be the CA corresponding to the local rule $\xi$.
Proposition~\ref{p:acp} can be applied to the CA $\Delta_\Upsilon\times G$ seen on the product alphabet $\Upsilon A^k$: it can be extended into a semifinite automaton $\tilde G$ on alphabet $\Upsilon A^k$, of anchor $0$, diameter $2$, and trace $\tau_{\tilde G}=\tau_{\Delta_\Upsilon\times G}$. In particular, its polytrace is $\ktau_{\tilde G}=\ktau_{\Delta_\Upsilon}\cup\ktau_G$.
Note that $\left(\xi\pow{k+l}\right)^{-1}(\Upsilon A^k)\subset\Upsilon A^k$.
Hence, Lemma~\ref{l:geltr0} allows to build a CA $F$ on alphabet $A$ such that $\tau_F=\ktau_{\tilde G}\cup\orb_\xi=\ktau_G\cup\orb_\xi$.
\qed

\proof[Proof of Proposition~\ref{p:polyctr}]
Let $G$ be a CA polytracing $\Sigma$.
 Let $\xi:A\to A$ be such that the deterministic subshift $\orb_\xi$ is not nilpotent and included in $\ktau_G$.
We can apply Lemma~\ref{l:bordtr0} to $G$ with $\Upsilon=\Upsilon^k_\xi$, which is $\frac{k+l}2$-freezing thanks to Lemma~\ref{l:gelks}, and $\xi\pow l$-invariant by construction, with $l=k+2$.
We obtain a CA $F$ such that $\tau_F=\ktau_G\cup\orb_\xi=\ktau_G$.
\qed

\end{document}